\documentclass[conference]{IEEEtran}
\IEEEoverridecommandlockouts
\usepackage{adjustbox}
\usepackage{amsmath,amssymb,amsfonts}
\usepackage{graphicx}
\usepackage{textcomp}
\usepackage{xcolor}
\usepackage{multicol}
\usepackage{multirow}
\usepackage{subcaption}
\usepackage{enumitem}
\usepackage{balance}
\usepackage{booktabs}
\usepackage{array}

\usepackage{pst-all}
\usepackage{pstricks-add}
\usepackage{float}

\usepackage{tikz}
\usepackage{tikzscale}
\usepackage{transparent}

\usepackage{algorithm}
\usepackage[noend]{algpseudocode}

\usepackage{comment}

\newcommand\blfootnote[1]{%
  \begingroup
  \renewcommand\thefootnote{}\footnote{#1}%
  \addtocounter{footnote}{-1}%
  \endgroup
}

\definecolor{purplelight}{HTML}{ead1dc}
\definecolor{purpledark}{HTML}{a64d79}

\definecolor{purple}{HTML}{733d87}

\definecolor{color1}{HTML}{e6194B}
\definecolor{color2}{HTML}{3cb44b}
\definecolor{color3}{HTML}{ffe119}
\definecolor{color4}{HTML}{4363d8}
\definecolor{color5}{HTML}{f58231}
\definecolor{color6}{HTML}{911eb4}
\definecolor{color7}{HTML}{42d4f4}
\definecolor{color8}{HTML}{f032e6}
\definecolor{color9}{HTML}{bfef45}
\definecolor{color10}{HTML}{fabebe}
\definecolor{color11}{HTML}{469990}
\definecolor{color12}{HTML}{e6beff}
\definecolor{color13}{HTML}{9A6324}
\definecolor{color14}{HTML}{a9a9a9}
\definecolor{color15}{HTML}{800000}
\def\BibTeX{{\rm B\kern-.05em{\sc i\kern-.025em b}\kern-.08em
    T\kern-.1667em\lower.7ex\hbox{E}\kern-.125emX}}

\begin{document}

\title{Automated Service Discovery for Social Internet-of-Things Systems}
\author{
\IEEEauthorblockN{Abdullah Khanfor\IEEEauthorrefmark{4},
  Hakim Ghazzai\IEEEauthorrefmark{4},
  Ye Yang\IEEEauthorrefmark{4}, Mohammad Rafiqul Haider\IEEEauthorrefmark{5},
  and Yehia Massoud\IEEEauthorrefmark{4}\\
%\IEEEauthorblockA{\IEEEauthorrefmark{1}College of Computer Science \& Information Systems, Najran University, Najran, Saudi Arabia}
\IEEEauthorblockA{\small \IEEEauthorrefmark{4}School of Systems \& Enterprises, Stevens Institute of Technology, Hoboken, NJ, USA\\
%Email: \{akhanfor, hghazzai, ye.yang, ymassoud\}@stevens.edu
}
\IEEEauthorblockA{\small \IEEEauthorrefmark{5}University of Alabama at Birmingham, AL, USA} %Email: mrhaider@uab.edu}
}\vspace{-0.4cm}}

\maketitle

\begin{abstract}
%The proliferation of massive Internet-of-things (IoT) has led to the emergence of many innovative applications such as mobile crowdsourcing. However, such ongoing technology is still rising a variety of challenges mainly related to the service discovery and navigation that allow IoT devices establish social relationships and cooperate together to accomplish certain tasks. 
In this paper, we propose to design an automated service discovery process to allow mobile crowdsourcing task requesters select a small set of devices out of a large-scale Internet-of-things (IoT) network to execute their tasks. To this end, we proceed by dividing the large-scale IoT network into several virtual communities whose members share strong social IoT relations. Two community detection algorithms, namely Louvain and order statistics local method (OSLOM) algorithms, are investigated and applied to a real-world IoT dataset to form non-overlapping and overlapping IoT devices groups. Afterwards, a natural language process (NLP)-based approach is executed to handle crowdsourcing textual requests and accordingly find the list of IoT devices capable of effectively accomplishing the tasks. This is performed by matching the NLP outputs, e.g., type of application, location, required trustworthiness level, with the different detected communities. The proposed approach effectively helps in automating and reducing the service discovery procedure and recruitment process for mobile crowdsourcing applications.
\end{abstract}

\blfootnote{\hrule
\vspace{0.2cm} This paper is accepted for publication in IEEE International Symposium on Circuits and Systems (ISCAS'20), Seville, Spain, Oct. 2020. \newline 2020 Personal use of this material is permitted. Permission from IEEE must be obtained for all other uses, in any current or future media, including reprinting/republishing this material for advertising or promotional purposes, creating new collective works, for resale or redistribution to servers or lists, or reuse of any copyrighted component of this work in other works.}%


\section{Introduction}

% What and why IoT and SIoT becomes more important
The rapid advancement in communication and sensor technologies have significantly raised the number of smart and connected devices. A forecast estimates that 15 smart devices will be owned and maintained by one person with a total of 500 billion devices connected to the Internet by 2030, according to CISCO~\cite{Cisco}. Typically, IoT is defined as a set of smart physical objects that are able to sense, report, and interact with the surrounding environment and hence, communicate and cooperate with each other and potentially establish ``social'' relations. Thus, the concept of social IoT (SIoT) has emerged to model the different relations that IoT objects can have and analyze their impacts on the IoT ecosystem and the application performance. The relationships between the different entities in IoT can be categorized into devices-to-devices, devices-to-people, and people-to-people~\cite{tan2010future, aono2016infrastructural}. Nevertheless, the heterogeneity and complexity of such network entities, structures, and social interactions pose challenges for practitioners to leverage the benefits of such a socially connected networks~\cite{bader2016front}.

% What are the types of objects and relationships that represents a network for IoT
One crucial aspect is to understand the different relations between these devices and employ them efficiently for better exploitation of the IoT network omnipresence and diversity \cite{hovhannisyan2019testing} \cite{khanfor2019application}. Indeed, individuals with social relationships and mobile computing capabilities are most likely to connect their devices with each other and initiate cooperative applications~\cite{atzori2012social}. Hence, in IoT systems, social relationships is not only limited to human interaction but can also be extended to machine-to-machine (M2M) interactions. According to~\cite{nitti2014network,afzal2019enabling,roopa2019social}, there are generally four types of relations that can be established among IoT devices: 1) parental-child relation, 2) co-location/co-work-based relation, 3) object ownership relation, and 4) social object relation.
%\textcolor{red}{Thus, we want to emphasize . Therefore, SIoT concept can be examined in this research which people and devices together forms a socio-technical network \cite{kranz2010things}. Another reason that the SIoT paradigm requires a free and dynamic establishment of relationships between the objects concerning their owners' assigned privileges \cite{nitti2012subjective}.}
% What are the issues and challenges in SIoT worth to investigate.
The existence of various social relationships in SIoT opens many research issues and thrust areas. For instance, it is mandatory to enhance the service discovery mechanisms in IoT networks in order to broaden the knowledge of human users about the different potential services. Moreover, when social relations are established, privacy and trustworthiness become important factors that impact the security aspects of the device's interaction. In this context and as an emerging application, mobile crowdsourcing can benefit from this broad SIoT system. In fact, by overlapping the mobile crowdsourcing and SIoT, mobile devices, e.g., smartphones or sensors, can be recruited as workers and be effectively exploited to meet the interests of requesters~\cite{Wan2019,AymenReporting}. %\textbf{One of the mobile crowdsourcing applications that can utilize the IoT system is the IoT inspector tool~\cite{huang2019iot}. This tool is used to monitor smart devices for the owners and allows to send public data about the status of their device as crowdsourcing application. Also, crowdsourcing and SIoT can be used to provide suitable devices (workers) in the IoT network to accomplish requested tasks. Therefore, the untapped benefits of computing and sensing capabilities of the crowd can be exploed in the SIoT context~\cite{gan2019crowdsocing}.}

%network navigability, trustworthiness management, and relationship management. 

%Service discovery concentrates on providing the services or information to human entities where the composition enables devices to interact with each other. Network navigability allows objects in the IoT network to establish different relations within constraints such as minimize the path length and permit for scaling the system. 

%Trustworthiness management focuses on the security aspects of the device's interaction and reliability. For relationship management, it concerns with privileges of different objects in the network and the update of these links and access between the objects. In our case, we use the mentioned relations in the literature between the devices and expended one more type of relation to accommodate the process of crowdsourcing a task in the IoT network.

% Crowdsourcing and IoT LAST PARAGRAPH ADDED <NEEDED TO BE REVISED CAREFULLY>

\begin{figure*}[htbp]
\centerline{\includegraphics[width=0.95\textwidth,height=6cm]{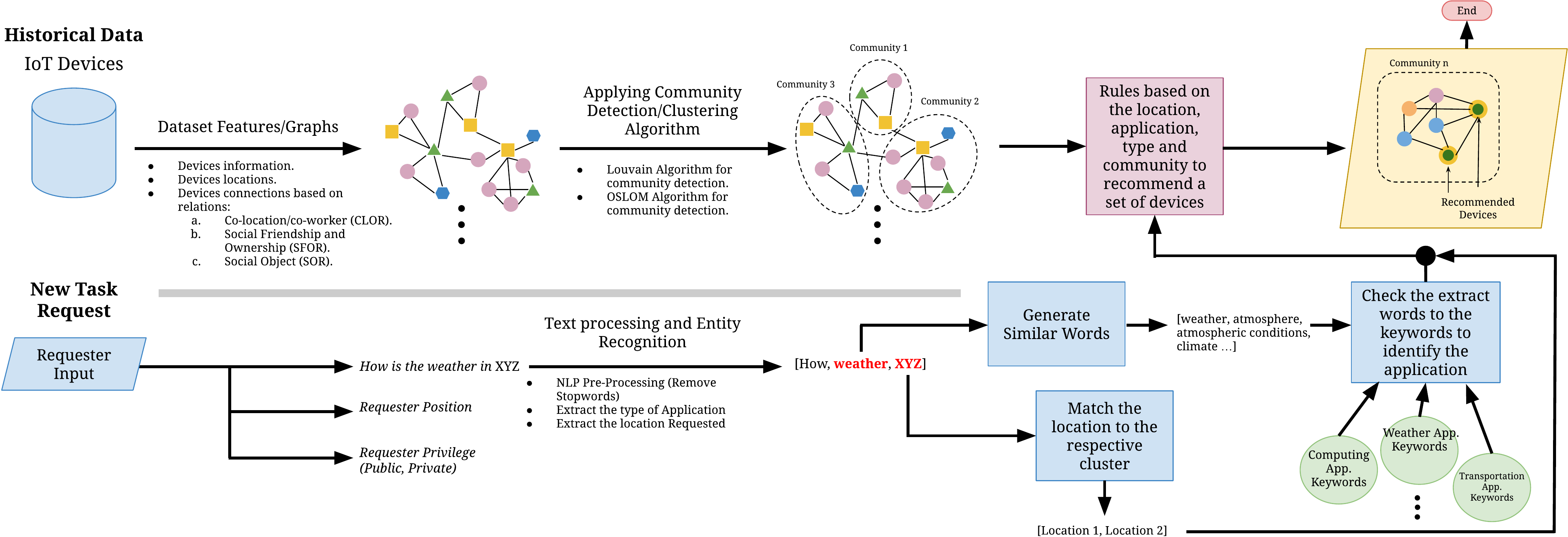}}
\caption{Proposed service discovery framework to map textual request to the identified communities in SIoT graph.}
\label{fig:framework}

\end{figure*}
In this paper, we design an automated service discovery process for mobile crowdsourcing applications while exploiting the benefits of SIoT infrastructure. The objective is to allow mobile crowdsourcing task requesters, usually human, to find a small set of devices out of a large-scale IoT network to execute their tasks. This will help in automating and reducing the service discovery procedure and recruitment process for many mobile crowdsourcing applications. In such a scenario, the crowdsourcing platform, instead for browsing its full dataset to find potential workers/devices, will extract information from the textual requests of the submitted tasks and accordingly find the list of IoT devices meeting the requirements of the crowdsourcers, e.g., application, location, trustworthiness level, etc. To this end, we divide the large-scale IoT network into several virtual communities whose members share strong social IoT relations. Two community detection algorithms, namely Louvain and OSLOM algorithms, are investigated and applied for this purpose to determine non-overlapping and overlapping IoT devices groups. Then, a NLP-based approach is executed to handle the textual requests and map them with different detected communities. The final list of IoT devices capable of responding the crowdsourcing task will correspond to ones obtained from the intersection of the identified communities. Selected simulation results are applied to a real-world IoT dataset and showcase some examples of the proposed service discovery process.

%In this paper, we are implementing different social relations between the devices' graphs. Also, we proposed a new relationship that we call it an Owners Friendship and Ownership relation (SFOR) that utilizes the social network between the owners and the previously proposed connections of the objects ownership relation. Based on that, we apply community detection algorithms to these relations to distinguish different groups of devices. Lastly, we showcase our framework by implementing a NLP workflow to handle a  request and then crowdsource it by recommending a shortlist of devices based on the communities identified from the first stage. By mapping the extracted information from the request to founded communities.

\section{Methodology}

In the proposed framework presented in Fig. \ref{fig:framework}, we focus on investigating the SIoT relations to help in the service discovery and crowdsourcing of mobile tasks. The framework includes two components: First, we split the IoT network into multiple communities sharing common characteristics using the community detection algorithm. Second, we propose a process to handle a request in natural language, e.g., ``What is the humidity level near the beach?''. The framework can extract valuable information from the textual request, such as the type of service and the location of the desired information. The requested service is then mapped to the classified communities of different types of relations such as location, friendship, and ownership from the first stage of the framework to determine a small set of IoT devices meeting the requirement of the crowdsourcing task requester.

\subsection{IoT Devices and Social Relationships}

We consider a large-scale IoT network located in a wide geographical area. The IoT objects have different types that vary from smartphones, smartwatches, weather sensors, and personal computers devices, etc. In order to build the social IoT network, we introduce the following social relationships:

$\bullet$ \textit{Co-location/co-work based relation (CLOR):}
This relationship is established based on the geospatial information of the devices. Hence, we assume that two IoT devices have a CLOR if the geographical distance separating them is less than a certain threshold. In other words, the devices are geographically co-located.

$\bullet$ \textit{Social friendship and ownership relation (SFOR):}
This relationship is established by considering the social relationship of the owners of the IoT devices. For example, two devices having the same owner are assumed to have a SFOR relationship. Another example could be the case of two devices owned by two different entities. But, if these owners are having any sort of social relationship (friends, cooperating), then they have some privilege to access their respective devices. Lastly, public devices are assumed to be owned by one entity, e.g., local authorities, that can be establishing a link with any entity in the network.

$\bullet$ \textit{Social object relationship (SOR):} The relationship is established when two IoT objects come into contact when operating. This relationship can be sporadic or continuous based on the needs of the devices or owner policies. For instance, we assume that if two devices have cooperated or communicated together in the past to complete a specific task, then they have a SOR relation. For example, they shared information together or cooperated in data routing, etc. This relationship is built based on the historical operation of the devices and their levels of interactions.

The three relationships presented earlier can be modeled by three different undirected and weighted graphs. The IoT devices constitute the vertices of the graphs while the edges represent the social IoT relationships between them. Some nodes may not be connected to one another for certain social relations. In that case, an edge will not be established. Moreover, the graph have no self-loop edges for the nodes. Finally, the weights on each edge indicate the strength of a relationship. In Section~\ref{sec:results}, we elaborate how social relations are established for real-world IoT system.

\subsection{Community Detection Algorithms} \label{sect:community}

In this section, we investigate community detection algorithms to split the IoT network on different virtual groups based on their social relationship. Flake et al.~\cite{flake2004graph} defined communities as a set of nodes that are connected with more edges compared to the rest of the network. There are different benefits of community detection applications, such as understanding the structure of a graph or improving information retrieval, which is one of our research aims. We apply two community detection algorithms to find non-overlapping and overlapping communities. Moreover, using communities as a method to reduce the search space instead of browsing the whole graph nodes by narrowing the search space to the respective community. As mentioned earlier, the objective is to simplify the service discovery process in large-scale IoT network and thus, the recruitment process in mobile crowdsourcing. To this end, we apply two community detection algorithms to find non-overlapping and overlapping communities.

%In this section, we investigate community detection algorithms to split the IoT network on different virtual groups based on their social relationship. Flake et al.~\cite{flake2004graph} defined communities as a set of nodes that are connected with more edges compared to the rest of the network. There are different benefits of community detection applications, such as understanding the structure of a graph or improving information retrieval, which is one of our research aims. Moreover, using communities as a method to reduce the search space instead of browsing the whole graph nodes by narrowing the search space to the respective community. As mentioned earlier, the objective is to simplify the service discovery process in large-scale IoT network and thus, the recruitment process in mobile crowdsourcing. To this end, we apply two community detection algorithms to find non-overlapping and overlapping communities.

\subsubsection{\textbf{Louvain Algorithm}}
Louvain method \cite{blondel2008fast} proposed in 2008 to detect non-overlapping, \textit{aka} disjoint, communities in a graph. The technique maximizes the modularity score for each community, where the modularity represents the quality of nodes assignment to communities by examining the density of edges with a set of nodes compared to how it would be connected in a random network. The Louvain method is a greedy optimization and one of the fastest community detection algorithms having a running time of $O(n \log n)$.

\subsubsection{\textbf{OSLOM Algorithm}} OSLOM is an overlapping clustering algorithm for graphs that conducts a local optimization of a fitness function by considering the statistical significance of clusters with respect to random fluctuations ~\cite{lancichinetti2011finding}. Even though the worst-case complexity of the algorithm is $O(n^2)$, the method shows a high accuracy in labeling the communities compared to similar methods in artificial network benchmarks.

\subsection{Task Assignment Workflow}
Once IoT virtual communities are identified, the mobile crowdsourcing platform can map textual requests to matching IoT groups and find a small set of devices capable of accomplishing the task. To do so, we adopt the process illustrated in Fig.~\ref{fig:framework} when a new task request is received. The text request is written in a natural language assumed to be fed by a human user. It includes generic information about the task such as type of requested service and task location. Moreover, with the submitted text, the request may include metadata such as the location of the requester and requester's trustworthiness level. The text is processed in the NLP pipeline to remove the stop words such as ``a'', ``an'', ``in'', ``the'', etc. Next, the words are tokenized to tag it with a part-of-speech (POS) labels. The POS labeling helps the framework extract the information needed to identify the type of application and the location the request needs to process the tasks. For example, the words that are proper nouns may indicate an area name. Other words are selected to generate a set of similar words. The generated words will be measured with the similarity of keywords list for each application such that each application have a set of pre-defined words. Then, we took the highest similarity score between the extracted words from the request and each application keywords. Finally, the framework will be able to locate the necessary information like application type, location, based on the highest similarity score. The new generated words will be compared with a list of pre-defined keywords and/or synonyms corresponding to each application.

In this study, we examine three application types about weather monitoring, transportation, and computation services. The application types can be expanded to other types. As an example, if the input text mentions the words weather, sunny, feel, rain, humidity, etc. then, the desired application will be about the weather, and therefore, it will require sensors with the capability of collecting weather-related data. Then, we extract the required location where the requester needs to check the weather and hence, the platform must identify the corresponding CLOR community. Indeed, using the community detection algorithm, the platform may determine multiple CLOR communities associated to a specific area on the map. Finally, the intersections of application type, CLOR, and SFOR communities will narrow down the list of devices and therefore provide a small set of devices that can process the request. %If the request is private, then it should check the requester's social network such that if they are friends or friends of a friend, then we can recommend the device for the requester.
\begin{comment}
\begin{figure}[t]
    \centering\resizebox{0.6\columnwidth}{!}{%
    \begin{tikzpicture}
        %ticks
    	\foreach \x in {0,...,9}
     		\draw (\x,1pt) -- (\x,-3pt) node[anchor=north] {};
    	\foreach \y in {0,...,8}
     		\draw (1pt,\y) -- (-3pt,\y) node[anchor=east] {}; 
        %custom ticks
        \node[text width=1] at (0.9,-0.5) {0};
        \node[text width=1] at (8.9,-0.5) {1};
        
        \node[text width=1] at (-0.5,1) {0};
        \node[text width=1] at (-0.5,8) {1};
        
        \node[anchor=south west,inner sep=0] at (1,1) {\includegraphics[width=0.9\columnwidth]{santander_maps_no_xy.png}};
        \draw[help lines, color=gray!70, dashed] (0,0) grid (9.9,8.9);
        \draw[->,thick] (0,0)--(10,0) node[right]{$x$};
        \draw[->,thick] (0,0)--(0,9) node[above]{$y$};
        \draw[draw=red, line width=2pt, fill opacity=0.1] (5,2) rectangle ++(2,5);
    \end{tikzpicture}
    %
    }\vspace{-0.2cm}
\caption{The geographical area of the city of Santander, Spain where 16216 IoT devices are deployed~\cite{marche2018dataset}.}
\label{fig:stand_map}\vspace{-0.5cm}
\end{figure}
\end{comment}
\section{Results \& Discussion}\label{sec:results}
\begin{figure*}[htbp]
        \centering
        \includegraphics[width=0.95\linewidth,height=5.9cm]{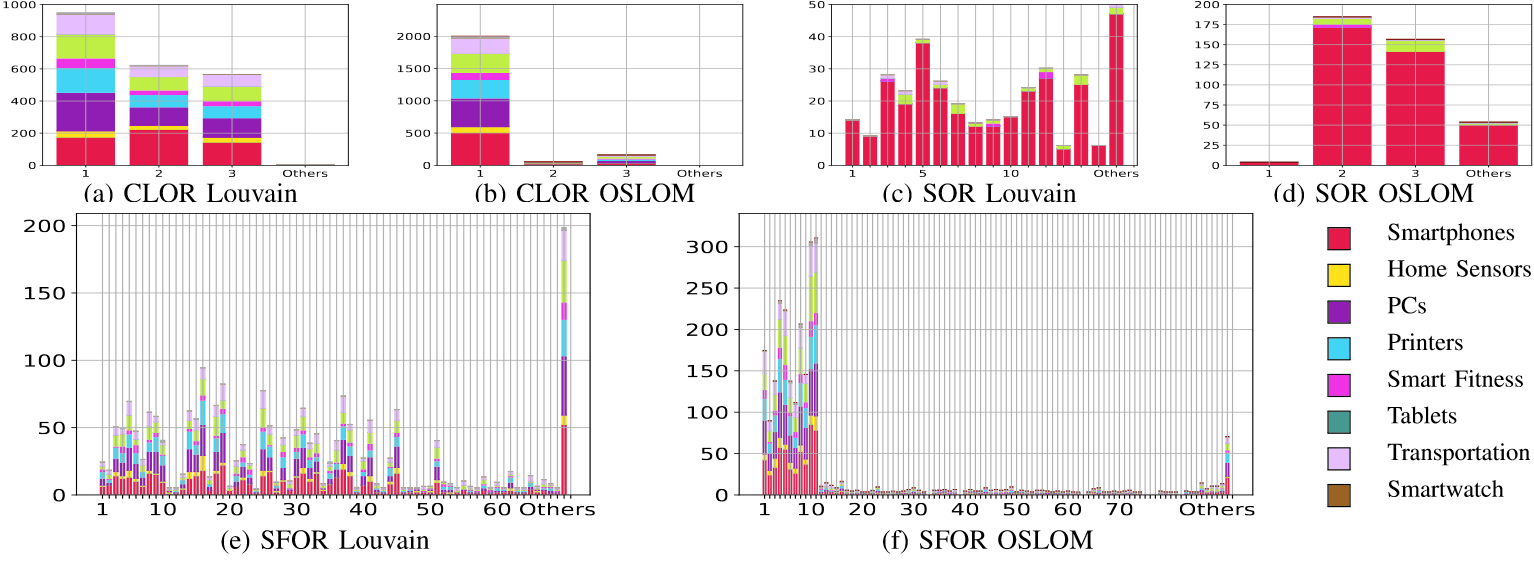}
        \vspace{-0.28cm}
        \caption{The frequency of devices in the respective community detected using Louvain and OSLOM algorithms.}\label{fig:bar_results}\vspace{-0.0cm}
\end{figure*}
To analyze our proposed framework, we use a real-world dataset of IoT devices that are stationed or projected in the city of Santander, Spain~\cite{marche2018dataset}. The total number of objects in the dataset is 16216 devices, of which 14600 are labeled as private users devices and 1616 are public devices. The devices vary from smartphones, smartwatches, weather sensors and personal computers devices. For clarity and tractability, we depict a small area where 2157 devices exist.% as indicated by the red box in Fig \ref{fig:stand_map}. 
We then run the community detection algorithms to obtain the virtual communities of the IoT network for the CLOR, SFOR, and SOR relations.

For the CLOR, to establish the links, we compute the edge weight $W_{ij}$ connecting to IoT devices $i$ and $j$ by measuring the normalized distance $d_{ij}$ as given in equations~\eqref{equ:dist_weights_norm}:
\begin{equation}\label{equ:dist_weights_norm}
W_{ij}= 1-\frac{d_{ij}}{d_{max}} \in [0,1],
\end{equation}
where $d_{max}$ is the longest distance that may separate two nodes in the map of interest. Then, we remove edges with a weight $W_{ij}$ less than the threshold $0.8$.

For the SFOR, we generate a random social network utilizing the Watts-Strogatz model~\cite{watts1998collective} since social relationships between owners are not available in the dataset. The model provides an undirected graph for all the nodes existing in the map representing their friendship levels. Hence, for SFOR, we create the edges between the devices based on the owners' friendship such that if one owner owns two devices, then an edge of weight equals to one is created. Afterwards, we connect all the devices owned by two friends with an edge of weight $0.75$. Finally, for a friend of a friend, we create an edge of weight equals to $\frac{1}{n_\text{hops}}$ between the devices where $n_{hops}$ is the number of hops connecting the owners in the social network.

%and set an average degree for the graph and a special parameter $0 \leq \beta \leq 1$. In this work, we set $N=4000$, which is the number of owners of the devices in the dataset. For the number of friends for each owner, we set $K=10$. And for adding a new edge between different nodes, we set $\beta=0.5$ in which represent small-world phenomena in the graph. 

Finally, for SOR, according to Marche et al.~\cite{marche2018dataset}, assumes that a SOR relationship is established between two objects if the objects have a three-time meeting of a duration not less than 30 minutes in total, and the interval between the two consecutive meetings is 6 hours.

\begin{figure}[t]
        \centering
        \begin{subfigure}[b]{\columnwidth}
        \centering
                \includegraphics[width=0.75\linewidth]{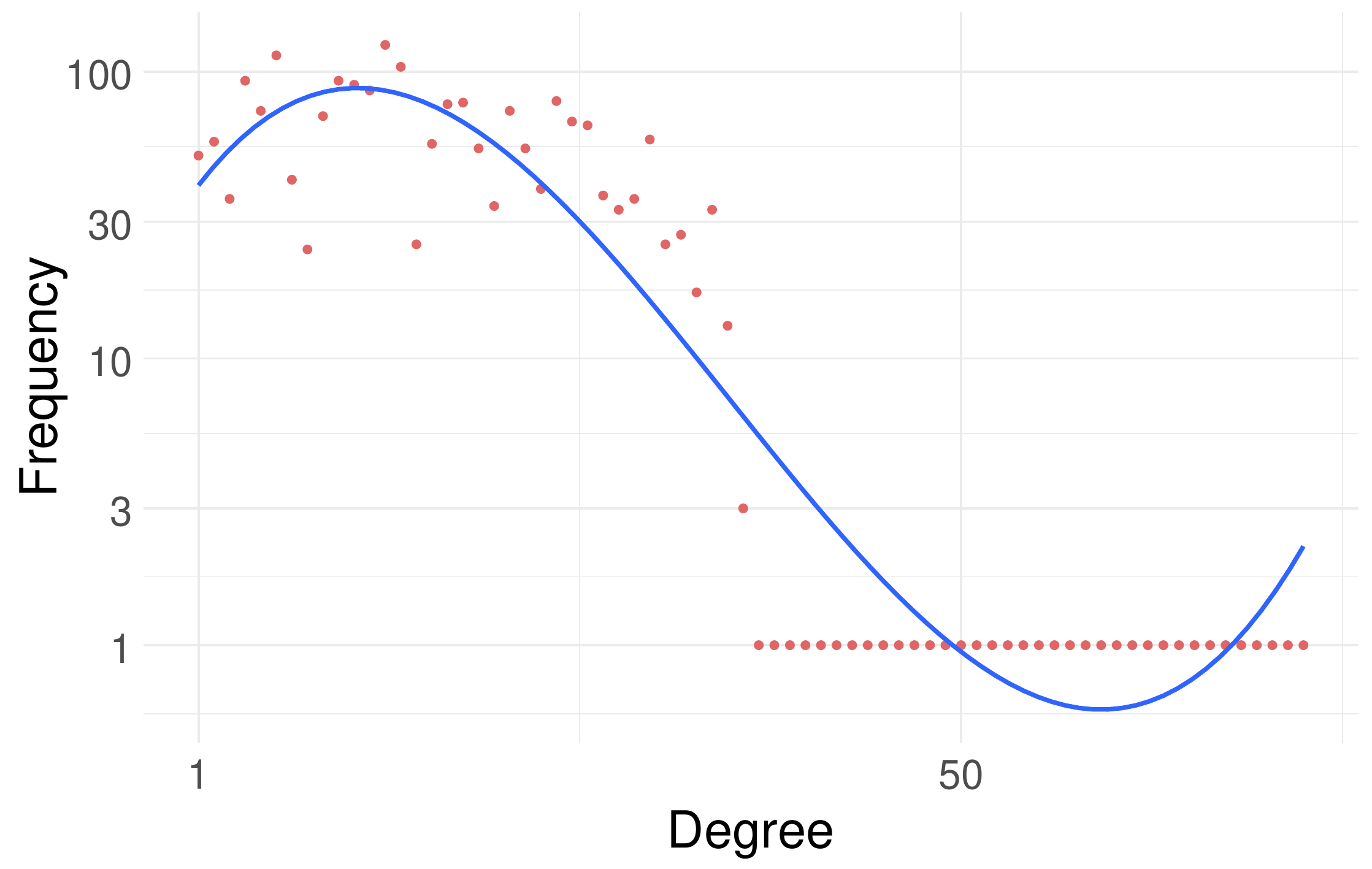}
                \caption{\scriptsize CLOR}
                \label{fig:dist_CLOR}\vspace{-0.2cm}
        \end{subfigure}%
        \hfill %%
        \centering
        \begin{subfigure}[b]{0.5\columnwidth}
                \includegraphics[width=\linewidth]{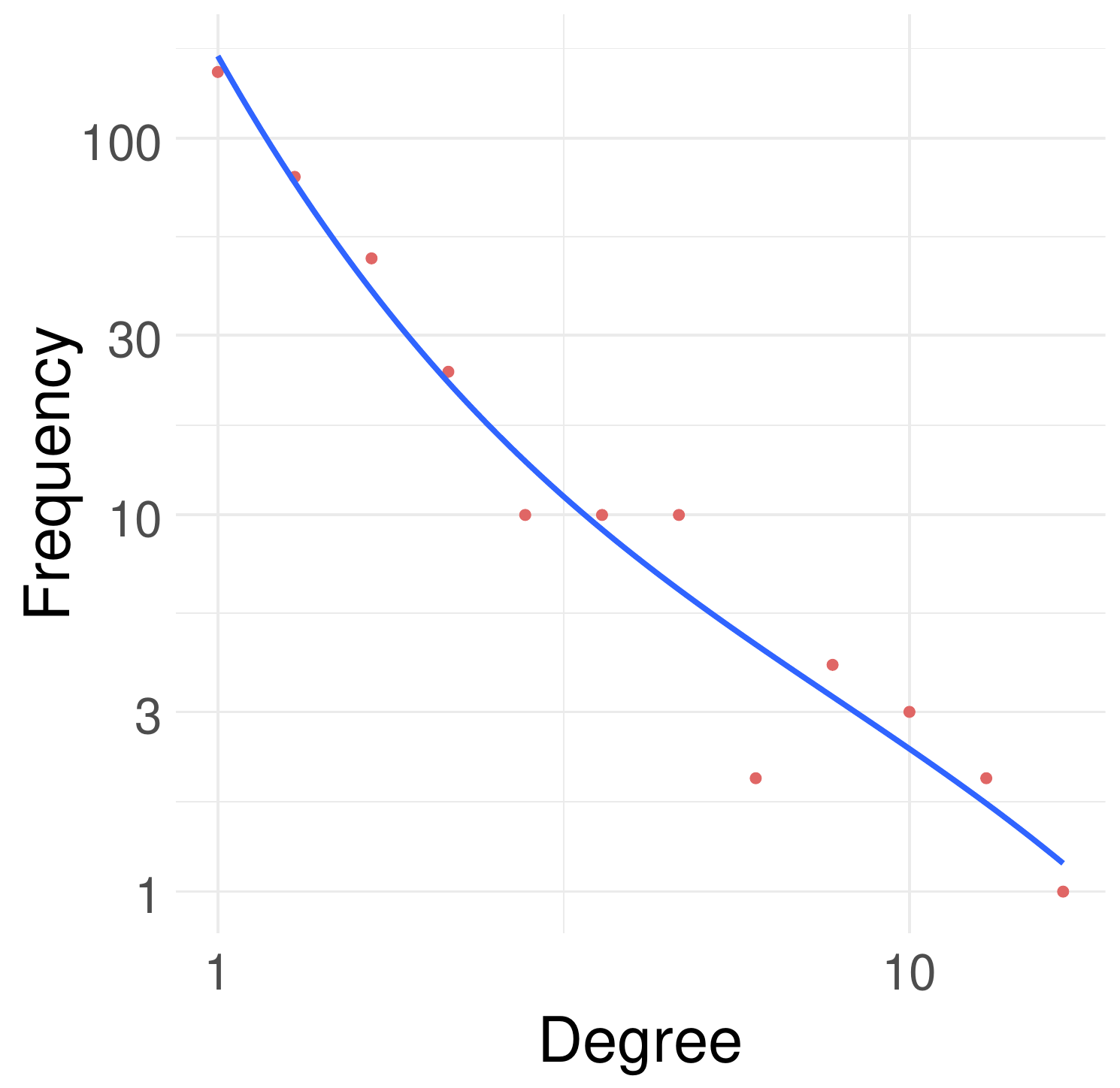}
                \caption{\scriptsize SOR}
                \label{fig:dist_SOR}\vspace{-0.2cm}
        \end{subfigure}%
        \begin{subfigure}[b]{0.5\columnwidth}
                \includegraphics[width=\linewidth]{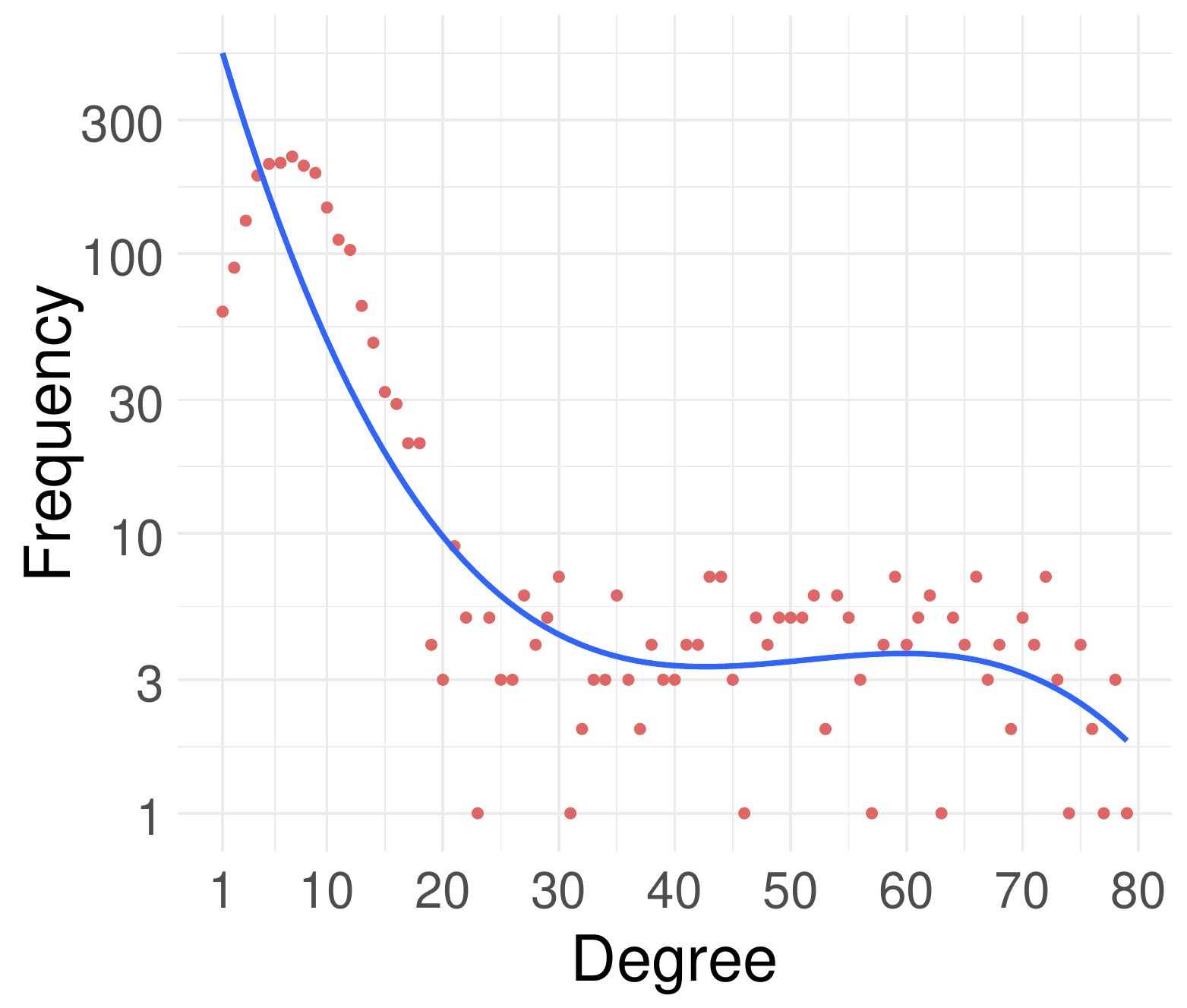}
                \caption{\scriptsize SFOR}
                \label{fig:dist_SFOR}\vspace{-0.2cm}
        \end{subfigure}%
        \hfill %%
    \caption{Degree distribution of graphs associated to each social IoT relations.}
    \label{fig:dist_degree}
\vspace{-0.2cm}
\end{figure}

In Fig.~\ref{fig:dist_degree}, we plot the network degree distribution for each social relation to identify the structure of the obtained graphs. The CLOR and SOR graphs exhibit a scale-free network where the number of nodes with a few degrees exceeds the number of nodes with a large number of connections. However, a small world graph is obtained for the SFOR relation since the network connections between the owners are based on the six degrees of the separation concept.

In Fig.~\ref{fig:bar_results}, we present the communities founded by Louvain and OSLOM algorithms. We group the communities having fewer than four devices under the ``Others'' label. For overlapping communities, the OSLOM algorithm can detect few but vast communities. However, the Louvain algorithm spreads the devices into communities having almost the same frequency of devices. For instance, with CLOR, the OSLM finds three communities, one of them with around 2000 devices while the rest have less than 300 devices. Louvain has three large communities having between 550 and 950 devices. Moreover, the average size of each community is larger in the CLOR relation since we are targeting a small area of the map. Hence, the devices are originally co-located. We also notice that the communities in SOR are only PCs, smartphones, and tablets; since it is based on a simulation that only established relationships among these devices. In summary, the Louvain algorithm provides reasonable virtual community distribution than OSLOM, which can be used to address mobile crowdsourcing requests.
\begin{figure}[tbp]
\centerline{\includegraphics[height=6.3cm]{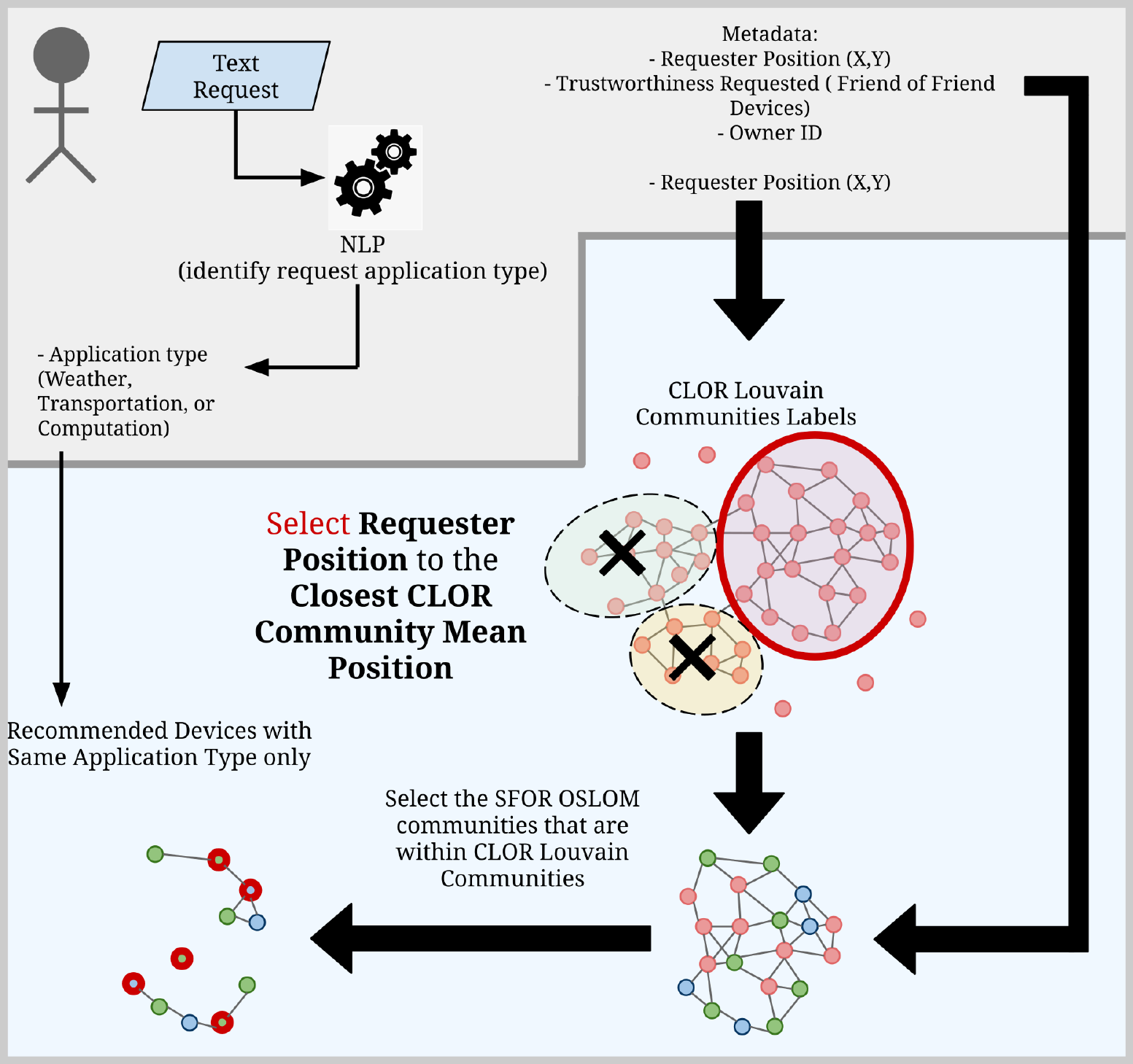}}\vspace{-0.1cm}
\caption{Framework system flow from the requester perspective.}
\label{fig:systemFlow}
\vspace{-0.3cm}
\end{figure}

We illustrate the workflow of the framework from the requester perspective in Fig. \ref{fig:systemFlow}. The gray area can be done on the client-side of the requester, such as smartphones or PCs; the blue part can be deployed on the server-side. The metadata from the requester will be used to recognize the position of the suitable CLOR community. The CLOR community will be labeled by the mean position of the devices of the community. Then, the framework will select the closest CLOR community to the requester position. Next, the SFOR communities within the CLOR will be selected based on the Friendship level requested. Finally, the devices will be recommended after comparing the request application type and the subsets of the CLOR Louvain community and SFOR OSLOM communities.

\section{Conclusion}

In this study, we proposed an automated framework to handle mobile crowdsourcing requests within a large-scale SIoT network. We first split the IoT network into multiple virtual communities grouping IoT devices sharing common characteristics modeled using the social relations. Two community detection algorithms with non-overlapping and overlapping communities are investigated. Then, with a NLP approach, we successfully capture the information from the textual request to match the with communities to find a small list of devices that would benefit the crowdsourcing process. %As a future work, we will investigate the real-time operation of the mobile crowdsourcing platform while considering the dynamic behavior of the social IoT network.

\balance
\bibliography{main}
\bibliographystyle{ieeetr}

\end{document}